\documentclass[conference]{IEEEtran}
\IEEEoverridecommandlockouts
\usepackage{balance}
\usepackage{cite}
\usepackage{amsmath,amssymb,amsfonts}
\usepackage{algorithmic}
\usepackage[ruled,vlined]{algorithm2e}
\usepackage{tcolorbox}
\usepackage{xcolor}
\usepackage{url}
\usepackage{bbold}

\definecolor{myblue}{HTML}{DAE8FC}
\definecolor{mygreen}{HTML}{D5E8D4}
\definecolor{myred}{HTML}{F8CECC}

\newcommand*\circled[1]{\tikz[baseline=(char.base)]{
    \node[shape=circle, draw, inner sep=0.2pt, minimum size=1.1em] (char) {\small #1};}}
\usepackage{tikz}
\usepackage{booktabs}
\usepackage{multirow}
\usepackage{graphicx}
\usepackage{textcomp}
\usepackage{xcolor}
\usepackage{comment}
\def\BibTeX{{\rm B\kern-.05em{\sc i\kern-.025em b}\kern-.08em
    T\kern-.1667em\lower.7ex\hbox{E}\kern-.125emX}}
\begin{document}

% \begin{comment}
\author{
\IEEEauthorblockN{Peihan Ye\IEEEauthorrefmark{1}, Alfreds Lapkovskis\IEEEauthorrefmark{1}, Alaa Saleh\IEEEauthorrefmark{2}, Qiyang Zhang\IEEEauthorrefmark{3}, and Praveen Kumar Donta\IEEEauthorrefmark{1} }

\IEEEauthorblockA{\IEEEauthorrefmark{1}\textit{Department of Computer Systems and Sciences}, \textit{Stockholm University}, Stockholm 164 25, Sweden} 
\IEEEauthorblockA{\IEEEauthorrefmark{2}\textit{Center for Applied Computing}, \textit{University of Oulu}, Oulu 90014, Finland} 
\IEEEauthorblockA{\IEEEauthorrefmark{3}\textit{Computer Science School}, \textit{Peking University}, Beijing 100876, China} 

% \texttt{{\color{red}peye5839@student.su.se}},
\texttt{peihanye5@gmail.com},
\texttt{alfreds.lapkovskis@dsv.su.se}, \texttt{alaa.saleh@oulu.fi}, \\ \texttt{qiyangzhang@pku.edu.cn},  \texttt{praveen@dsv.su.se}
}
% \end{comment}
% \author{Anonymous Authors}
\title{NeSy-Edge: Neuro-Symbolic Trustworthy Self-Healing in the Computing Continuum}
\maketitle

\begin{abstract}
The computational demands of modern AI services are increasingly shifting execution beyond centralized clouds toward a computing continuum spanning edge and end devices. However, the scale, heterogeneity, and cross-layer dependencies of these environments make resilience difficult to maintain. Existing fault-management methods are often too static, fragmented, or heavy to support timely self-healing, especially under noisy logs and edge resource constraints. To address these limitations, this paper presents \texttt{NeSy-Edge}, a neuro-symbolic framework for trustworthy self-healing in the computing continuum. The framework follows an edge-first design, where a resource-constrained edge node performs local perception and reasoning, while a cloud model is invoked only at the final diagnosis stage. Specifically, \texttt{NeSy-Edge} converts raw runtime logs into structured event representations, builds a prior-constrained sparse symbolic causal graph, and integrates causal evidence with historical troubleshooting knowledge for root-cause analysis and recovery recommendation. We evaluate our work on representative Loghub datasets under multiple levels of semantic noise, considering parsing quality, causal reasoning, end-to-end diagnosis, and edge-side resource usage. The results show that \texttt{NeSy-Edge} remains robust even at the highest noise level, achieving up to 75\% root-cause analysis accuracy and 65\% end-to-end accuracy while operating within about 1500 MB of local memory.
\end{abstract}


\section{Introduction}\label{sec:intro}
\IEEEPARstart{A}{rtificial} intelligence (AI) is becoming part of everyday systems and services; it needs more computing power, faster response, and better resource management across the computing continuum (CC). Increasing operating costs force organizations to rethink how and where computation for these services should run. Recent research suggests that computing is increasingly distributed across cloud platforms, edge nodes, and end devices, creating a CC that meets the demands of modern AI applications \cite{dustdar2022distributed}. While CC is designed with several goals, including improved response time, local data use, and resource flexibility \cite{sedlak2025adaptive}, it also introduces new system challenges. These environments are often large, heterogeneous, and tightly interconnected, with dependencies across computational resources, often resource-constrained, data flows, Internet connections, energy demands, cooling systems, and orchestration mechanisms \cite{sedlak2026service}. Because of these interdependencies, a fault in one layer or device can quickly cascade across other parts of the continuum, resulting in service disruptions. For CC, resilience therefore becomes a basic requirement rather than an optional feature. These systems must sustain operation under faults, adapt to changing conditions, limit disruption, and recover quickly without significant loss of performance or efficiency \cite{chen2026resilient}.

In practice, achieving resilience in the CC remains challenging because of its highly dynamic and uncertain nature \cite{donta2025resilientdesignactive}. Resource availability can vary over time, network quality may fluctuate, devices may join or leave unexpectedly, and workloads may be dynamically orchestrated across cloud, edge, and end nodes. In the literature, some works \cite{11126953,Poveda2020FedGuard} have explored fault tolerance and recovery in the CC, but these works often rely on static rules, isolated monitoring, or centralized control, which limit their ability to explain faults and coordinate timely recovery under highly dynamic conditions. The problem becomes challenging in resource-constrained settings, where limited resources restrict the use of heavyweight fault-management methods. Moreover, resilience in the CC is not only about replacing failed components, but also about maintaining service continuity and making recovery decisions that meet system and application goals. This calls for mechanisms that can observe system state at runtime and reason about failures, cross-layer dependencies, and feasible recovery actions. Such capability is critical in the CC, where faults may be local and transient or may cascade across orchestration layers, data flows, and shared infrastructure. Therefore, effective resilience in the CC must extend beyond basic fault tolerance and enable adaptive, coordinated self-healing under dynamic operating conditions.

The above-mentioned limitations highlight the need for a novel self-healing approach for the CC. In such environments, resilience requires not only fault detection but also the ability to interpret runtime observations, anticipate possible failures, and make timely recovery decisions. Moreover, such a solution should not be limited to resource-rich platforms but should also operate on resource-constrained edge devices, handle noisy and heterogeneous logs, and support recovery across multiple layers of the continuum. 

\textbf{Contribution.} We propose \texttt{NeSy-Edge}, a three-layer neuro-symbolic framework for trustworthy self-healing in the CC. Our framework addresses the aforementioned challenges by transforming raw runtime logs into root-cause analysis (RCA) reports and actionable recovery decisions, thereby reducing unnecessary data transfer, supporting more consistent fault diagnosis, and enabling context-aware recovery across the continuum. The design of the three \texttt{NeSy-Edge} layers is summarized as follows:
\begin{itemize}
    \item \textit{Perception layer:} converts noisy logs into structured event representations while keeping
    parsing operations local and lightweight. Here, raw runtime logs are preprocessed and routed through a three-tier mechanism comprising exact symbolic cache matching, retrieval-assisted semantic matching over a historical knowledge base, and a lightweight local small-language-model (SLM) fallback for unseen patterns. 
    \item \textit{Reasoning layer:} constructs a prior-constrained sparse symbolic causal graph from structured event streams. The method combines sliding-window event aggregation with historical causal priors to infer sparse causal dependencies and suppress unlikely relations.
    \item \textit{Action layer:} combines causal evidence with retrieved troubleshooting knowledge for RCA and recovery recommendations. It also includes a selective inference mechanism to reduce cloud dependence when local evidence is sufficient for deterministic decision-making.
\end{itemize}

We assess the proposed framework under various scenarios using multiple datasets and evaluation metrics. To the best of our knowledge, closely aligned end-to-end baselines for this problem setting remain limited. Therefore, we focus on layer-wise evaluation by comparing each component with its task-specific baselines. 

The rest of this paper is organized as follows. Section~\ref{sec:relatedworks} presents a review of the most recent and relevant literature. 
Section~\ref{sec:sysmodel} describes the system model.
Section~\ref{sec:proposed} describes the proposed \texttt{NeSy-Edge} framework design. Section~\ref{sec:results} presents and discusses the simulation environment, datasets used, and numerical results and performance analysis. Finally, Section~\ref{sec:concl} concludes the paper.

\section{Related Work}\label{sec:relatedworks}
Recently, several works focused on log sequence modeling for fault detection and RCA to achieve fault tolerance and resilience, most recent and suitable works are discussed below. 

Recent works use log analysis by integrating language models to improve semantic understanding and accuracy in detecting failures in computer systems. For example, Hong et al.~\cite{11229682} combine small models for efficiency with LLMs for accuracy through diversified sampling, rule-based LLM triggering, and dynamic template merging. Similarly, Zhang et al.~\cite{11216353} use LLM-based token classification with a semantic perception module and prefix-tree structures for scalable template matching. Also, self-correcting in-context learning, asynchronous scheduling, and controversy-based sampling are other strategies used to further enhance adaptability and performance in the literature~\cite{11025900,11113127}. Additionally, semantic representations of logs to improve anomaly detection performance gain further attention while identifying the root cause of failures. In particular, GPT has been used to semantically enhance log templates, followed by Sentence-Bidirectional Encoder Representations from Transformers (BERT)-based embedding extraction and attention-based fusion of the original and enhanced representations. This hybrid semantic modeling is further complemented by Transformer-based anomaly detection \cite{11025902}. Further, an unsupervised Transformer-based approach has been adopted \cite{10930057} using the BERT Masked Language Model to detect anomalies in log parameters. Yang et al.~\cite{11292191} integrate FastText embeddings, TF-IDF weighting, and a robust principal component analysis (PCA) model with adaptive loss to improve efficiency and resilience under dynamic log patterns.

Zhang et al.~\cite{10888670} propose a RAG approach combining log summarization, sample augmentation, and Chain-of-Thought prompting to diagnose failures directly from raw logs. Xu et al.~\cite{11334373} integrate token-efficient preprocessing, structured diagnostic prompting, and RAG-based reasoning with tool-calling to not only identify root causes but also execute automated remediation actions using historical knowledge. Mamanpoosh et al.~\cite{11297608} adopt a divide-and-conquer strategy through hierarchical summarization, where large volumes of unstructured logs are segmented, summarized with domain-specific prompts, and aggregated into coherent diagnostic reports, particularly for IoT systems. Alongside these LLM-driven reasoning approaches, Huang et al.~\cite{11036790} integrate data dependency graph construction with a dynamic log-tracking system that combines regular expressions and LLM-based semantic analysis for real-time fault localization. The modular system proposed in \cite{11172898} integrates log collection, preprocessing, and LLM-based semantic analysis for automated classification and anomaly detection. Similarly, a hybrid multi-layered framework was proposed in  \cite{11252439} that combines rule-based classification,sentence-embedding clustering, supervised learning (KNN), and an LLM for automated log classification and the early detection of critical issues in massive systems. 

In the literature, some works do not adapt LM-based models for fault tolerance or resilience. For example, a federated learning framework integrates self-attention with Bidirectional LSTM to capture temporal and contextual patterns in distributed environments~\cite{11406610}. Similarly, a deep learning-based approach for fault detection and diagnosis has been proposed in \cite{11173039}, which integrates log parsing with LSTM-based sequence modeling and hierarchical fault classification, enabling the capture of temporal dependencies and anomaly detection across system levels. Kodela et al.~\cite{11209721} propose a hybrid multimodal architecture that fuses a Bi-LSTM of log sequences with CNN-based performance metrics, followed by a Random Forest classifier and SHAP-based explainability to enable interpretable RCA. Evolutionary approaches such as genetic algorithms have also been used to develop efficient graph-based parsers with low computational cost \cite{11073703}. Moving toward autonomous systems, Saleh et al.~\cite{saleh2026bioinspired} propose a bio-inspired architecture employing LM-driven multi-agent layers for fault detection, causal reasoning, and adaptive recovery, enabling continuous learning and self-managing behavior in complex environments.

Unlike prior studies that mainly address isolated stages such as log parsing, anomaly detection, LLM-based diagnosis, or recovery orchestration, our work presents an integrated edge-first neuro-symbolic framework for trustworthy self-healing in the CC. \texttt{NeSy-Edge} is distinguished by its end-to-end three-layer design, which combines lightweight local log perception, prior-constrained symbolic causal reasoning, and causal-evidence-guided RCA and recovery recommendation within a single pipeline. In contrast to approaches that rely primarily on centralized or purely neural inference, our framework explicitly incorporates symbolic priors, selective cloud invocation, and resource-aware local execution, making it better suited to noisy, heterogeneous, and resource-constrained continuum environments. 

\section{System Model}\label{sec:sysmodel}

The CC spans one or $C$ cloud data centers $\mathcal{C}=\{c_i\}_{i=1}^C$, and $E$ other heterogeneous nodes $\mathcal{E}=\{e_j\}_{j=1}^{E}$, such as Internet-of-Things (IoT) devices, edge, fog, etc. Thus, CC can be represented as a graph $G_{CC}=(\mathcal{V},\mathcal{R})$, where $\mathcal{V}=\mathcal{C}\cup\mathcal{E}$ are nodes, and $\mathcal{R}\subseteq \mathcal{V}\times\mathcal{V}$ communication relationships between them. In this paper, we focus on resource-constrained edge nodes; however, our framework is generalizable to other types of devices in both $\mathcal{E}$ and $\mathcal{C}$. 

At each time step $t$, each node $v\in\mathcal{V}$ executes its own set of tasks (or services) $\mathcal{T}_t(v)$, and can experience zero or more faults (or incidents) $\mathcal{F}_t(v)$. Each fault $f$ can span a continuous sequence of time steps, which we denote as an incident \emph{window} $w_f=\{t,\dots,t+N-1\}$, where $N$ is the window length. During an incident window, the monitored service generates a raw log stream $L_{raw}^f=\{l_1,\dots,l_M\}$, where $M$ denotes the number of logs generated for the fault.

The objective of the proposed \texttt{NeSy-Edge} framework is to ensure self-healing of nodes $v\in\mathcal{V}$ from failures $f\in\mathcal{F}_t(v)$. To achieve that, the framework performs a sequence of steps to transform raw logs $L_{raw}^f$ into RCA reports $R_{final}^f=(r^*,a^*)$, where $r^*$ denotes the diagnosed root cause of $f$ and $a^*$ is the recommended corrective action. The produced RCA report is used both to address the current fault and to improve future RCA generation.

The following section describes the proposed framework in detail. For better readability, we omit the $f$ superscript from $L_{raw}$, $R_{final}$, and other variables' notation.

\section{Proposed Framework} \label{sec:proposed}
\begin{figure}[!t]
    \centering
    \includegraphics[width=0.495\textwidth]{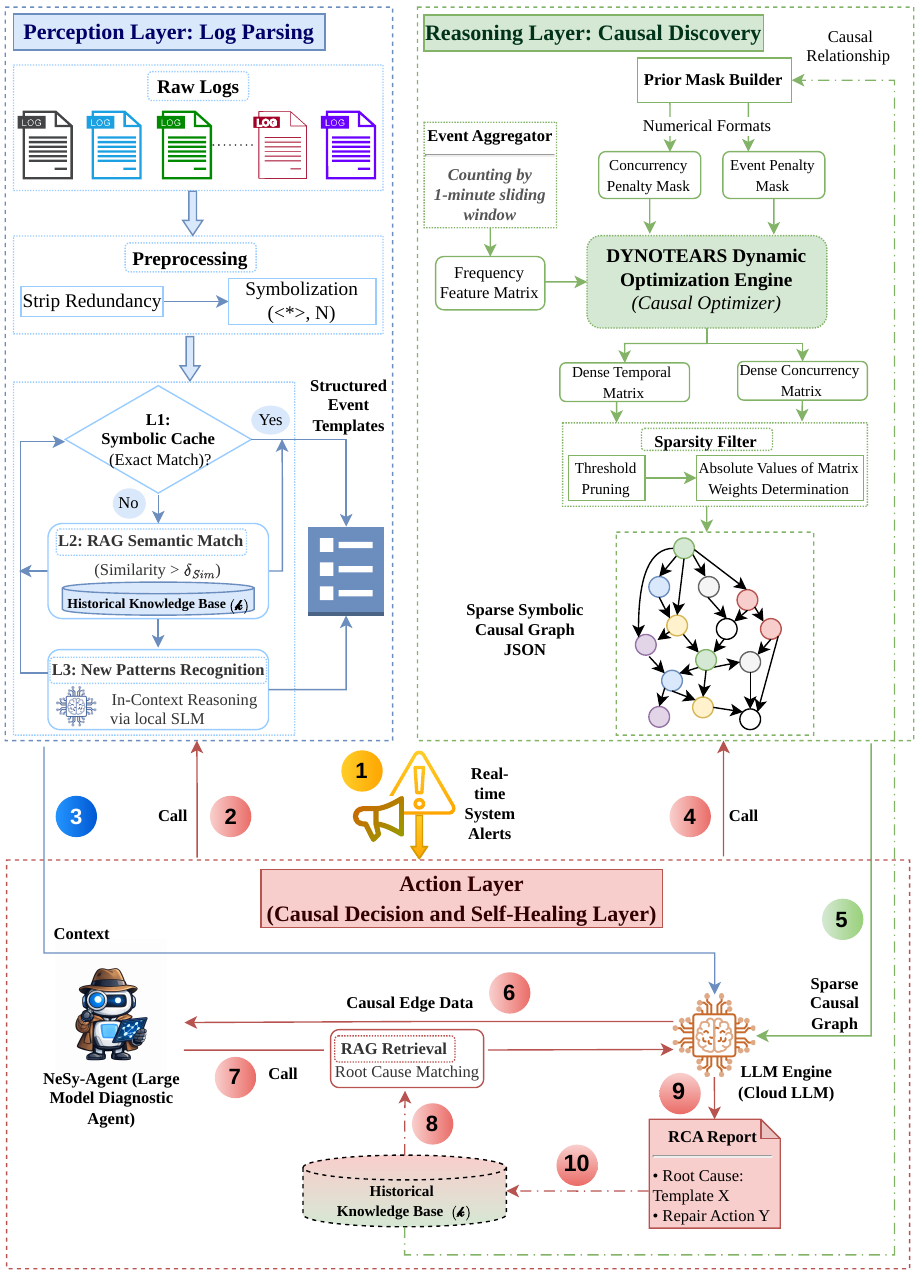}
    \caption{The overall architecture and execution workflow of \texttt{NeSy-Edge}}
    \label{fig1}
\end{figure}

As illustrated in Fig.~\ref{fig1}, \texttt{NeSy-Edge} is organized into three layers: the perception layer, the reasoning layer, and the action layer. The perception layer parses raw logs into structured event templates, the reasoning layer constructs a prior-constrained sparse symbolic causal graph, and the action layer combines causal evidence with historical knowledge to generate the final RCA report and repair action. The main execution steps are labeled \circled{1} to \circled{10} in Fig.~\ref{fig1} for reference.

The workflow is event-driven. When anomalies at edge nodes trigger real-time system alerts \circled{1}, the action layer orchestrates the subsequent local perception and reasoning steps and, when needed, the final diagnosis step for RCA and self-healing.

\subsection{Perception Layer}\label{subsec:perception}

At step \circled{2}, the action layer invokes the perception layer to convert the raw log stream $L_{raw}$ into a set of structured event templates $T_{struct}$. The overall parsing procedure is summarized in Algorithm~\ref{alg:perception}. The perception layer maintains a small local symbolic cache for exact reuse of validated templates, while the shared historical knowledge base $\mathcal{K}$ stores a broader set of reference entries for semantic retrieval and later downstream reuse.

First, each raw log is preprocessed by removing redundant information and replacing dynamic fields, such as IP addresses and timestamps, with generic placeholders (e.g., \verb|<*>|). The original timestamps are retained separately for the subsequent sliding-window aggregation in the reasoning layer. The processed log, denoted by $l'$, is then passed through a three-tier router that balances efficiency and generalization.

The first tier, L1 cache matching, performs an exact lookup against the local symbolic cache. If no exact match is found, the second tier performs retrieval-assisted matching by querying the historical knowledge base $\mathcal{K}$ and applying cosine similarity. Let $\mathbf{e}_{l'}$ denote the embedding of the processed log $l'$, and let $\mathbf{e}_{k}$ denote the embedding of a candidate reference entry $k \in \mathcal{K}$. A semantic match is accepted when the maximum similarity exceeds a predefined threshold:
\begin{equation}
\max_{k \in \mathcal{K}} \mathrm{Sim}(\mathbf{e}_{l'}, \mathbf{e}_{k})=\max_{k \in \mathcal{K}}
\frac{\mathbf{e}_{l'}^\top \mathbf{e}_{k}}
{\|\mathbf{e}_{l'}\|_2 \|\mathbf{e}_{k}\|_2}
\ge \delta_{Sim},
\label{eq:cosine}
\end{equation}
where $\delta_{Sim}$ is a configurable similarity threshold. If this condition is satisfied, the matched reference template is reused; otherwise, the log is treated as an unseen pattern and passed to the third tier, namely a local lightweight LLM fallback, which abstracts a new template through in-context reasoning.

To support continuous adaptation, a template generated by the local fallback is inserted into the L1 cache immediately for local reuse, while incorporation into $\mathcal{K}$ is performed asynchronously after validation. After parsing is complete, the perception layer returns $T_{struct}$ at step \circled{3} to the action layer as structured context for the subsequent reasoning stage.

\begin{algorithm}[t]
\caption{Three-Tier Log Parsing Router of Perception Layer}
\label{alg:perception}
\SetKwInput{Input}{Input}
\SetKwInput{Output}{Output}
\SetKwComment{Comment}{$\triangleright$ }{}
\Input{Raw log stream $L_{raw}$, Local L1 cache $\text{L1\_Cache}$, Historical knowledge base $\mathcal{K}$}
\Output{Structured event templates $T_{struct}$}
$T_{struct} \leftarrow \emptyset$\;
\ForEach{log $l \in L_{raw}$}{
    $l' \leftarrow \text{Preprocess}(l)$ \Comment*[r]{Normalize dynamic fields}
    \uIf{$\text{ExactMatch}(l', \text{L1\_Cache})$}{
        $t \leftarrow \text{Fetch}(l')$ \Comment*[r]{L1: exact cache match}
    }
    \uElseIf{$\max_{k \in \mathcal{K}} \mathrm{Sim}(\mathbf{e}_{l'}, \mathbf{e}_{k}) \ge \delta_{Sim}$}{
        $t \leftarrow \text{FetchSemantic}(l')$ \Comment*[r]{L2: retrieval-assisted match}
    }
    \Else{
        \begin{tcolorbox}[colback=white, colframe=myblue, coltitle=black, title= Local LLM Prompt, width=0.82\linewidth]
        - TASK $\rightarrow$ Recognize new log patterns.\\
        - Execute In-Context Reasoning to abstract template.
        \end{tcolorbox}
        $t \leftarrow \text{LLM\_Generate}(l')$ \Comment*[r]{L3: local LLM fallback}
        $\text{UpdateCache}(\text{L1\_Cache}, t)$ \Comment*[r]{Immediate local update}
        $\text{UpdateKB}(\mathcal{K}, t)$ \Comment*[r]{After validation}
    }
    $T_{struct} \leftarrow T_{struct} \cup \{t\}$\;
}
\Return{$T_{struct}$}\;
\end{algorithm}

\subsection{Reasoning Layer}\label{subsec:reasoning}
At step \circled{4}, the action layer invokes the reasoning layer to construct a sparse symbolic causal graph from the structured event templates $T_{struct}$. The overall reasoning procedure is summarized in Algorithm~\ref{alg:reasoning}.

First, an event aggregator groups the templates in $T_{struct}$ using a fixed 1-minute sliding window and counts their occurrences to form a frequency feature matrix $X \in \mathbb{R}^{m \times d}$, where $m$ is the number of time windows and $d$ is the number of distinct event types. 
So that each element $X_{u,j}$ records the number of occurrences of event type $j$ in window $u$.
The same event ordering is then used consistently also by $W_{mask}$, $A_{mask}$, $W$, and $A$, which are defined below.

Next, a prior mask builder encodes historical causal knowledge from the historical knowledge base $\mathcal{K}$ as two penalty masks for optimization: (i) an intra-slice penalty mask $W_{mask} \in \mathbb{R}^{d \times d}$ for dependencies within the same time slice, whose elements are computed as
\begin{equation}
    W_{mask,ij} =
        \begin{cases}
        c_W^{prior}, & \text{if } (i,j)\in P_W \\
        c_W^{rev}, & \text{if } (j,i)\in P_W \land (i,j)\notin P_W \\
        c_W^{bg}, & \text{otherwise},
        \end{cases}
\end{equation}
where $P_W$ are directed event pairs that are historically supported within the same time slice;
and (ii) an inter-slice penalty mask $A_{mask} \in \mathbb{R}^{d \times d}$ for lagged dependencies across adjacent time slices, whose elements are computed as
\begin{equation}
    A_{mask,ij} =
        \begin{cases}
        c_A^{prior}, & \text{if } (i,j)\in P_A  \\
        c_A^{rev}, & \text{if } (j,i)\in P_A \land (i,j)\notin P_A \\
        c_A^{bg}, & \text{otherwise},
        \end{cases}
\end{equation}
where $P_A$ are directed event pairs that are historically supported across adjacent time slices.
The penalty values $c\in[0,\infty)$ reflect how strongly a candidate relation should be encouraged or discouraged. In particular, a small penalty $c^{prior}_\cdot$ is assigned to event-template pairs that are supported by historical evidence, a background penalty $c^{bg}_\cdot$ is assigned to unknown relations, and a large penalty $c^{rev}_\cdot$ is assigned to implausible reverse-direction relations.

We formulate causal discovery by extending the DYNOTEARS algorithm\cite{pamfil2020dynotears}. The symbolic prior knowledge is explicitly injected into the continuous optimization problem through element-wise product ($\circ$):

\begin{equation}
\min_{W, A} \mathcal{L}(X, W, A) + \lambda_W \| W \circ W_{mask} \|_1 + \lambda_A \| A \circ A_{mask} \|_1
\label{eq:dynotears_opt}
\end{equation}
\begin{equation}
\text{s.t. } h(W) = \text{tr}(e^{W \circ W}) - d = 0,
\label{eq:acyclic}
\end{equation}
where $\mathcal{L}(X, W, A)$ is the least-squares data-fitting term, $W$ is the dense weighted adjacency matrix for intra-slice dependencies within the same time slice, $A$ is the dense weighted adjacency matrix for inter-slice dependencies across adjacent time slices, and $\lambda_W$ and $\lambda_A$ are regularization weights that control the strength of the prior penalties. The acyclicity constraint in Eq.~(\ref{eq:acyclic}) is imposed only on $W$, because inter-slice edges in $A$ connect earlier slices to later slices and therefore do not create same-slice directed cycles \cite{pamfil2020dynotears}.

\begin{algorithm}[t]
\caption{Prior-Constrained Causal Discovery}
\label{alg:reasoning}
\SetKwInput{Input}{Input}
\SetKwInput{Output}{Output}
\SetKwComment{Comment}{$\triangleright$ }{}
\Input{Structured event templates $T_{struct}$, Historical knowledge base $\mathcal{K}$}
\Output{Sparse symbolic causal graph $G_{causal}$}
$W_{mask}, A_{mask} \leftarrow \text{BuildPriorMask}(\mathcal{K})$\;
Initialize dense matrices $W$ (intra-slice), $A$ (inter-slice)\;
\textbf{Optimize objective function:} Solve Eq.~(\ref{eq:dynotears_opt}) subject to Eq.~(\ref{eq:acyclic})\;
\ForEach{weight $w_{ij} \in W \cup A$}{
    \If{$|w_{ij}| < \theta_{prune}$}{
        $w_{ij} \leftarrow 0$ \Comment*[r]{Threshold pruning}
    }
}
$G_{causal} \leftarrow \text{ConstructGraph}(W, A)$\;
\Return{$G_{causal}$}\;
\end{algorithm}

The masking terms in Eq.~(\ref{eq:dynotears_opt}) increase the optimization cost for implausible edges while preserving historically supported relations. After optimization, edges with magnitudes below the pruning threshold $\theta_{prune}$ are removed to improve graph compactness. Prior to serialization, the framework may slightly reduce the weights of weak edges near a threshold, and retain the direction with higher prior knowledge support when two directions still have similar plausibility after pruning. This improves the graph's usability. The reasoning layer then returns the resulting sparse symbolic causal graph $G_{causal}$ at step \circled{5} to the action layer as structured causal evidence, preserving event-template labels, directed relations, and edge weights for downstream causal navigation.

\subsection{Action Layer}\label{subsec:action}
The action layer generates the final RCA report from the sparse symbolic causal graph and the historical knowledge base. This stage is executed by the \textit{NeSy-Agent}, which first extracts compact causal evidence from the graph, then retrieves matched historical troubleshooting cases, and finally determines the root cause and repair action either locally or, when ambiguity remains, through a cloud LLM. The overall diagnostic procedure is summarized in Algorithm~\ref{alg:action}.

At step \circled{6}, the \textit{NeSy-Agent} invokes a causal navigator to traverse $G_{causal}$ and extract directed causal evidence, denoted by $E_{causal}$. This evidence summarizes the graph region most relevant to the current incident, including candidate root events, their key upstream relations, and the highest-weight directed paths. Rather than passing the full graph directly downstream, the agent uses this compact evidence as the structured basis for diagnosis. When cloud inference is needed, the agent applies a causal-constrained prompt that limits the model to relations explicitly supported by $E_{causal}$, instead of introducing additional causal links.

At steps \circled{7} and \circled{8}, the agent performs RAG retrieval over the historical troubleshooting knowledge base $\mathcal{K}$ to obtain matched troubleshooting knowledge, denoted by $K_{match}$. The retrieval query is formed from the event labels and candidate root events contained in $E_{causal}$, so that the returned cases are aligned with the current causal pattern rather than with surface similarity alone. The retrieved cases are then combined with $E_{causal}$ as grounded context for the final decision.

If the graph-derived evidence and retrieved cases already converge to the same diagnosis result without remaining ambiguity, the \textit{NeSy-Agent} outputs the RCA report locally and bypasses the cloud LLM. Otherwise, at step \circled{9}, the cloud LLM receives the combined context $E_{causal} \oplus K_{match}$ and produces the final RCA report by selecting a root cause and a corresponding repair action supported by the causal evidence and historical cases. Finally, only validated diagnosis-action pairs are updated into $\mathcal{K}$ to support future retrieval, and the final RCA report is returned at step \circled{10}.

\begin{algorithm}[t]
\caption{NeSy-Agent Diagnostic Workflow}
\label{alg:action}
\SetKwInput{Input}{Input}
\SetKwInput{Output}{Output}
\SetKwComment{Comment}{$\triangleright$ }{}
\Input{Sparse causal graph $G_{causal}$, Historical troubleshooting knowledge base $\mathcal{K}$}
\Output{Final RCA report $R_{final}=(r^\star, a^\star)$}

\begin{tcolorbox}[colback=white, colframe=mygreen, coltitle=black, title= NeSy-Agent Prompt, width=0.92\linewidth]
- Recognize only causal facts explicitly established in the graph.
\end{tcolorbox}

$E_{causal} \leftarrow \text{CausalNavigator}(G_{causal})$ \circled{6}\;
$K_{match} \leftarrow \text{RetrieveCases}(E_{causal}, \mathcal{K})$ \circled{7,8}\;

\eIf{$\text{DeterministicMatch}(E_{causal}, K_{match})$}{
    $R_{final} \leftarrow \text{LocalDecision}(E_{causal}, K_{match})$\;
}{
    \begin{tcolorbox}[colback=white, colframe=myred, coltitle=black, title= LLM Synthesis, width=0.92\linewidth]
    - INPUT: $E_{causal} \oplus K_{match}$\\
    - DECIDE: Select the supported root cause and repair action.
    \end{tcolorbox}
    $R_{final} \leftarrow \mathcal{M}_{LLM}(E_{causal} \oplus K_{match})$ \circled{9}\;
}
$\text{UpdateKnowledgeBase}(\mathcal{K}, R_{final})$ \Comment*[r]{After validation}
\Return{$R_{final}$ \circled{10}}\;
\end{algorithm}

\section{Results and Discussion}\label{sec:results}

We implemented the proposed \texttt{NeSy-Edge} in Python and integrated all methods into a unified framework, where case construction, noise injection, and metric computation were executed within a single deterministic evaluation pipeline. Our implementations are available in the GitHub repository\footnote{\url{https://anonymous.4open.science/r/NeSy-Edge-D794}}.

\subsection{Experimental Setup}

Although the framework was evaluated on a standard workstation (MacBook Pro with Apple M4, 10-core CPU, 24\,GB RAM), we explicitly designed the pipeline to emulate a resource-constrained edge device, such as a typical IoT gateway. Accordingly, we strictly isolated the local monitoring path, comprising the exact cache, symbolic rules, retrieval stage, and the Qwen3-0.6B model~\cite{yang2025qwen3}. This local path was executed under resource-constrained edge settings and monitored throughout all accepted runs. In practice, the monitored local process remained below 1,500\,MB peak resident set size (RSS), indicating that local execution stayed well within the $2$\,GB limit on edge devices without requiring additional computational resources. To keep our framework executable within the available local resources, the experiments were run in a chunked, resumable manner. The heavyweight cloud LLM, DeepSeek-V3.2~\cite{deepseek2025v32}, is explicitly excluded from the local monitoring loop and is invoked only via API during the final agentic diagnostic stage. In our experiments, we inject controlled semantic perturbations across six noise levels $\{0.0, 0.2, 0.4, 0.6, 0.8, 1.0\}$. We adapt the perturbation strategy to the distinct semantic structures of different system logs. For short control-plane events, the injector rewrites synonymous tokens (e.g., swapping ``GET'' with ``FETCH''). For storage and heterogeneous application logs, it perturbs parser-visible variables, anchors, and status terms. This mechanism simulates realistic edge-telemetry ambiguities and tests whether the framework remains robust, i.e., whether it maintains performance as semantic noise increases under surface-form variations.  Additional configuration specifications are summarized in Table~\ref{tab:setup_summary}.

\begin{table}[t]
\centering
\caption{Summary of Experimental Configurations}
\label{tab:setup_summary}
\begin{tabular}{ll} 
\toprule
\textbf{Component} & \textbf{Specification / Configuration} \\
\midrule
\textbf{Host Environment} & Apple M4 (10 Cores), 24\,GB RAM, macOS \\
\textbf{Local Edge Model} & Qwen3-0.6B\\
\textbf{Cloud Agent Model}& DeepSeek-V3.2 (API only) \\
\textbf{Generation Temperature} & $0.0$ (Strict determinism for diagnosis) \\
\textbf{Noise Levels}     & $\{0.0, 0.2, 0.4, 0.6, 0.8, 1.0\}$ \\
\textbf{Local Execution Budget} & 2 CPU threads, 2048\,MB target memory \\
\textbf{Observed Local Peak RSS} & $< 1500$\,MB\\
\bottomrule
\end{tabular}
\end{table}

\subsection{Datasets}
Our experiments are conducted on three representative datasets from the public Loghub collection\cite{loghub}. These datasets were considered to cover three complementary log characteristics: structured templates, control-plane events with high ambiguity, and heterogeneous long-window failures. For the perception layer, a small disjoint subset of clean logs is reserved as reference data for exact matching and retrieval, while the remaining cases are used for evaluation under noisy conditions.

\begin{itemize}
    \item \textbf{HDFS:} Represents storage-service logs with recurring block-management patterns. For the perception benchmark, we use 160 clean reference logs and 2,000 evaluation cases. For the reasoning and action benchmarks, we use 46 benchmark cases and 50 complete incident windows, respectively.

    \item \textbf{OpenStack:} Contains short control-plane messages that are semantically similar and therefore prone to ambiguity under noise. For the perception benchmark, we use 192 clean reference logs and 1600 evaluation cases. For the reasoning and action benchmarks, we use 97 benchmark cases and 50 complete incident windows, respectively.
    
    \item \textbf{Hadoop:} Represents complex and more heterogeneous resource failures. Its logs are longer and structurally more diverse than those of HDFS and OpenStack. For the perception benchmark, we use 160 clean reference logs and 1,400 evaluation cases. For the reasoning and action benchmarks, we use 47 benchmark cases and 44 complete incident windows, respectively.
    
\end{itemize}
Across all datasets, the perception layer contains 5,000 evaluation cases, which expand to 90{,}000 method-by-noise result rows under three methods and six noise levels. The reasoning layer evaluates 190 test cases in total, while the action layer contains 144 complete incident windows, yielding approximately 2,600 autonomous diagnostic runs.

\subsection{Baselines}
To isolate the contribution of each component in the proposed three-layer framework, we compare \texttt{NeSy-Edge} with representative task-specific baselines at the perception, reasoning, and action layers.

\begin{itemize}
    \item \textbf{Perception:} We compare against \textbf{Drain} \cite{he2017drain}, representing traditional pure pattern-matching symbolic parsers, and a \textbf{Direct SLM} baseline (implemented via Qwen3-0.6B \cite{yang2025qwen3}), representing pure neural template generation without symbolic caching and retrieval routing.

    \item \textbf{Reasoning:} We evaluate structural graph quality against three distinct paradigms: the prior-free original \textbf{DYNOTEARS}\cite{pamfil2020dynotears} representing continuous score-based optimization, the classical \textbf{PC algorithm} \cite{spirtes2000causation} representing constraint-based conditional independence, and a standard \textbf{Pearson} correlation graph representing naive statistical co-occurrence.
    
    \item \textbf{Action:} For downstream diagnosis, we compare against a \textbf{Vanilla LLM}\cite{chen2024rcacopilot} baseline to test base generative capabilities directly on the selected alert and incident context, and a \textbf{RAG-only}\cite{xu2025logsage} baseline representing the use of standard retrieved historical references without graph-derived causal evidence.

\end{itemize}

\subsection{Numerical results}
This section presents the numerical results of \texttt{NeSy-Edge} compared with baseline methods across the HDFS, OpenStack, and Hadoop datasets with varying noise levels, for the perception, reasoning, and action layers.

\subsubsection{Evaluations on Perception Layer}
In this layer, we evaluate parsing accuracy and inference latency across three datasets at varying levels of noise.

\begin{figure}[!t]
    \centering
    \includegraphics[width=0.495\textwidth]{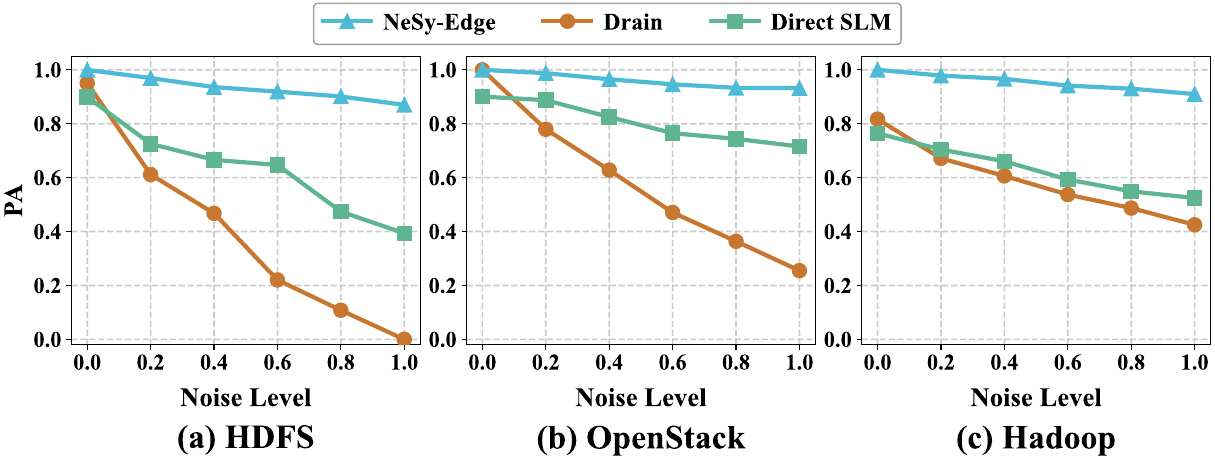}
    \caption{Parsing Accuracy under increasing semantic noise.}
    \label{fig:rq1_pa}
\end{figure}

\textbf{Parsing accuracy (PA)} measures the fraction of alert cases whose predicted template exactly matches the ground-truth template after normalization.
PA is computed as
\begin{equation}
    \mathrm{PA}=\frac{1}{N}\sum_{i=1}^N \mathbb{1}\left[\mathrm{Norm}(\hat{t}_i)=\mathrm{Norm}(t_i)\right],
\end{equation}
where $N$ is the number of evaluated logs, $\hat{t}_i$ is the predicted template, $t_i$ is the ground truth template, and $\mathrm{Norm}(\cdot)$ is the normalization function.
Higher PA indicates stronger robustness in recovering the correct template from noisy logs. Fig.~\ref{fig:rq1_pa} shows a consistent performance across all three datasets over six noise levels. \texttt{NeSy-Edge} achieves the highest PA throughout over Drain, whereas Drain performs well only when the input format remains stable. Also, the Direct SLM baseline shows intermediate robustness but lower exact-match consistency than \texttt{NeSy-Edge}. This result indicates that the hybrid routing design of \texttt{NeSy-Edge} is more reliable than either a purely rule-based or a purely neural parser alone.

For example, on HDFS (Fig.~\ref{fig:rq1_pa}(a)), Drain drops sharply from $0.95$ to $0$, showing that its tree-based matching becomes unreliable once the original token structure is heavily altered. The Direct SLM degrades more gradually, from $0.90$ to $0.39$, remaining comparatively stable up to the middle noise range but dropping more clearly at higher noise levels, whereas Drain exhibits a much earlier cliff-like failure. In contrast, \texttt{NeSy-Edge} remains substantially more robust, decreasing only from $0.99$ to $0.87$. This shows that combining cache matching, retrieval-assisted matching, and local LLM fallback remains effective even when the input wording is strongly changed.

On OpenStack (Fig.~\ref{fig:rq1_pa}(b)), \texttt{NeSy-Edge} again performs best, while Drain still declines rapidly and the Direct SLM degrades much more gradually. Drain decreases from $1$ to $0.25$, while the Direct SLM declines from $0.90$ to $0.71$. This contrast suggests that once lexical anchors are disturbed, exact tree-based matching becomes brittle, whereas direct neural abstraction remains relatively tolerant to the same perturbation. By comparison, \texttt{NeSy-Edge} maintains PA above $0.93$ even at the highest noise level. This behavior is consistent with the structure of OpenStack control-plane logs, where many messages retain reusable patterns despite semantic noise.

The Hadoop results further highlight the benefit of the proposed design on more heterogeneous logs, as shown in Fig.~\ref{fig:rq1_pa}(c). Compared with HDFS and OpenStack, Hadoop contains longer and more diverse log patterns, which makes exact recovery harder for purely rule-based matching. Accordingly, Drain starts lower at $0.81$ and declines to $0.42$, while the Direct SLM decreases from $0.76$ to $0.52$. The two baselines follow more similar and relatively gradual trajectories than on HDFS, suggesting that on longer and more heterogeneous templates, both exact rule matching and direct generation are constrained by the same structural complexity. \texttt{NeSy-Edge} again remains clearly superior, dropping only from $1$ to $0.91$ while increasing the noisy levels. This suggests that on structurally diverse logs, the advantage of \texttt{NeSy-Edge} lies not only in exact cache reuse, but also in its ability to exploit retrieval-assisted matching before invoking the local neural fallback.

From these observations, we noticed that purely symbolic parsing is efficient but increasingly fragile under semantic noise, while purely neural parsing is more flexible but less reliable for exact template recovery. By combining both, \texttt{NeSy-Edge} provides the most robust recovery behavior across all evaluated settings.

\begin{figure}[!t]
    \centering
    \includegraphics[width=0.495\textwidth]{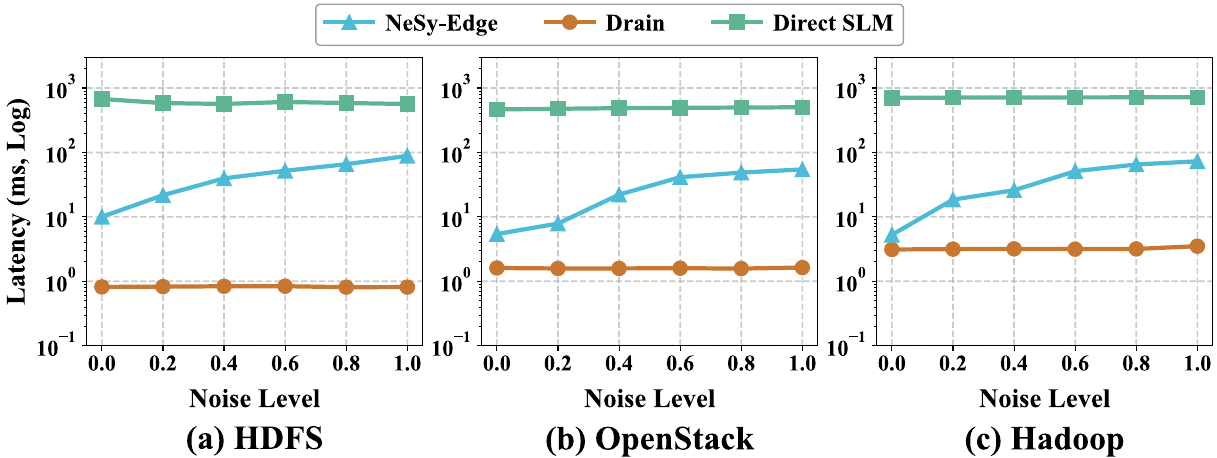}
    \caption{Inference latency under increasing semantic noise.}
    \label{fig:rq1_latency}
\end{figure}

% \subsubsection{Inference Latency (Perception Layer)}
\textbf{Inference latency} measures the wall-clock time required to process a single alert. Under the emulated edge budget, low cost is important, but so is the ability to spend additional computation only when the input becomes difficult. From Fig.~\ref{fig:rq1_latency}, Drain is the fastest method across all datasets and all noise levels. Its latency remains nearly flat because the parsing process is fully local and rule-based. At the other extreme, the Direct SLM is the slowest baseline, since every input triggers a full local generation step regardless of difficulty.
\texttt{NeSy-Edge} exhibits a different cost profile. Instead of assigning the same computation to every alert, it uses a routing strategy in which easier cases are handled through cache matching or retrieval-assisted matching, while cases with new patterns are sent to the local LLM fallback. As a result, its latency remains far below the Direct SLM baseline across all settings, gradually increasing as the noise level rises.

The dataset-level differences are also informative. Fig.~\ref{fig:rq1_latency}(a) shows performance variations on HDFS. Here, the average latency of \texttt{NeSy-Edge} increases from $10.03$\,ms at noise $0.0$ to $88.55$\,ms at noise $1.0$. This larger increase is consistent with the sharper loss of stable input structure in HDFS under strong noise, leading more cases to leave the fast-matching routes and enter fallback processing. On OpenStack (from Fig.~\ref{fig:rq1_latency}(b)), the increase is milder, from $5.41$\,ms to $54.47$\,ms, indicating that many noisy cases can still be resolved without invoking the local LLM. As shown in Fig.~\ref{fig:rq1_latency}(3) about the performance on Hadoop datasets, \texttt{NeSy-Edge} rises from $5.26$\,ms to $72.91$\,ms. Although its latency at noise $0.0$ is close to that of OpenStack, it rises to a higher cost under strong noise, consistent with the greater heterogeneity of Hadoop logs. 

Taken together with the PA results, the latency results confirm the practical value of the proposed perception layer design. \texttt{NeSy-Edge} is not intended to outperform a purely symbolic parser in raw speed. Instead, it reduces the cost of neural invocations while preserving strong robustness to noise. This is the behavior required for an edge perception module: most alerts are handled through lightweight local matching, and additional computation is used only when the input becomes ambiguous.

\subsubsection{Evaluations on Reasoning Layer}
The reasoning layer evaluates the structural quality of the causal graphs generated by different reasoning methods. Here, we evaluate two performance metrics, i.e.,  \textbf{Sparsity} denotes the number of edges in the final graph, where lower values indicate a more compact and interpretable structure, and \textbf{average root-cause rank (AvgRank)} measures how highly the ground-truth root-cause relation is placed among the ranked candidate edges under the task-aligned evaluator. AvgRank is computed as
\begin{equation}
    \mathrm{AvgRank}=\frac{1}{N}\sum_{i=1}^N r_i,
\end{equation}
where $M$ is the number of benchmark cases, and $r_i$ is the rank position assigned to the ground truth root cause relation in case $i$. If the correct relation is not found in the graph, we assign to $r_i$ the number of graph edges $+1$ as a miss penalty.
Lower AvgRank values and values closer to $1.0$ are better. Both Sparsity and AvgRank should be considered jointly, since a useful causal graph should remain compact while still placing the correct root-cause relation near the top of the candidate list.

\begin{figure}[!t]
    \centering
    \includegraphics[width=0.495\textwidth]{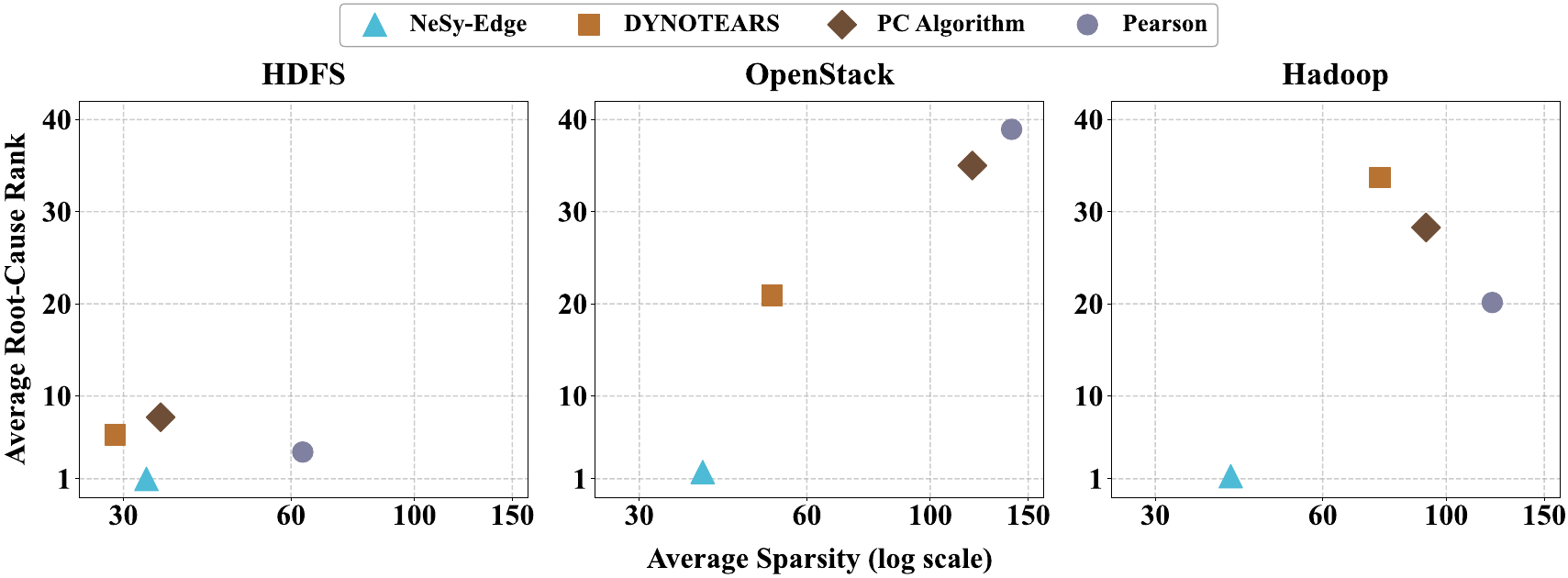}
    \caption{Structural quality of reasoning-layer causal graphs across baselines.}
    \label{fig:rq2_pareto}
\end{figure}

Fig.~\ref{fig:rq2_pareto} shows the tradeoff between these two objectives. Across all three datasets, \texttt{NeSy-Edge} is consistently closest to the desirable lower-left region, indicating a better balance between graph compactness and root-cause localization than the baselines. Pearson and PC tend to produce denser or less directionally useful graphs, whereas prior-free DYNOTEARS can remain relatively compact while still retaining spurious dependencies. By contrast, \texttt{NeSy-Edge} uses symbolic priors to guide edge selection and ranking, which yields graphs that are both compact and useful for downstream diagnosis.

The detailed values are given in Table~\ref{tab:rq2_data}. On HDFS, the benchmark is relatively concentrated, so some baselines can still correctly identify the cause at a moderately good position. For example, Pearson reaches an AvgRank of $3.91$, but does so with a much denser graph of Sparsity $63.0$. DYNOTEARS yields the smallest graph on this dataset with Sparsity $29$, but its AvgRank remains worse at $5.78$. \texttt{NeSy-Edge} slightly increases the edge count to $33$, yet improves the rank to $1.0$, showing that the gain comes from retaining a more useful set of edges rather than simply removing more of them.

The advantage of symbolic priors is evident on OpenStack. Because this dataset contains short and highly similar control-plane messages, co-occurrence can easily lead to confusion. Under this setting, Pearson and PC both become much denser and rank the correct cause poorly, while DYNOTEARS reduces graph size but still leaves the correct relation far from the top of the ranked candidates. In comparison, \texttt{NeSy-Edge} achieves both the most compact graph and the best ranking result, with Sparsity $39$ and AvgRank $1.74$. This shows that, on OpenStack, the prior-constrained design is valuable not only for reducing graph size but also for suppressing implausible dependencies.

Finally, the Hadoop results further confirm the robustness of the proposed reasoning design in a more heterogeneous setting. In Hadoop datasets, \texttt{NeSy-Edge} provides the best trade-off, with Sparsity $41$ and AvgRank $1.28$. The remaining methods all produce substantially denser graphs and weaker ranking results, indicating that symbolic priors remain effective when the event space is broader and the candidate search space is harder to control.

Overall, these results support the reasoning layer's design goal. \texttt{NeSy-Edge} improves the structural usefulness of the causal graph by placing the correct root-cause relation closer to the top while keeping the graph compact and interpretable. This provides stronger, structured evidence for the downstream action layer.

\begin{table}[htbp]
\centering
\caption{Structural quality of causal graphs across baselines}
\label{tab:rq2_data}
\resizebox{\linewidth}{!}{
\begin{tabular}{llcccc}
\toprule
\textbf{Dataset} & \textbf{Metric} & \textbf{NeSy-Edge} & \textbf{DYNOTEARS} & \textbf{PC} & \textbf{Pearson} \\
\midrule
\multirow{2}{*}{\textbf{HDFS}}
& Sparsity & \textbf{33} & 29 & 35 & 63 \\
& Avg. Rank & \textbf{1.00} & 5.78 & 7.69 & 3.91 \\
\midrule
\multirow{2}{*}{\textbf{OpenStack}}
& Sparsity & \textbf{39} & 52 & 119 & 140 \\
& Avg. Rank & \textbf{1.74} & 20.92 & 35.01 & 38.93 \\
\midrule
\multirow{2}{*}{\textbf{Hadoop}}
& Sparsity & \textbf{41} & 76 & 92 & 121 \\
& Avg. Rank & \textbf{1.28} & 33.71 & 28.28 & 20.14 \\
\bottomrule
\end{tabular}
}
\end{table}
% --------------------------------

\subsubsection{Evaluations on Action Layer}
\begin{figure}[!t]
    \centering
    \includegraphics[width=0.495\textwidth]{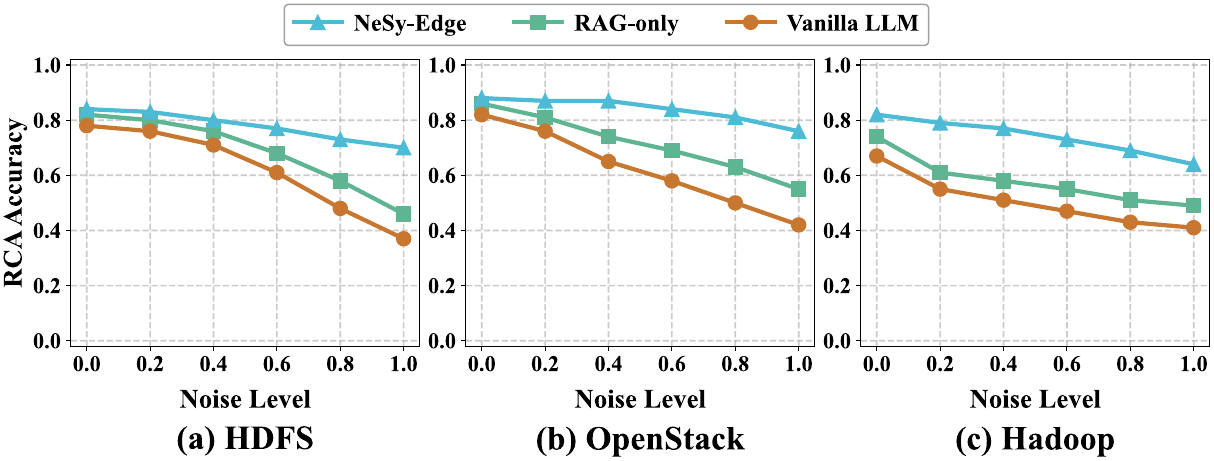}
    \caption{RCA accuracy under increasing semantic noise.}
    \label{fig:rq3_rca}
\end{figure}

Action Layer evaluates downstream diagnosis on complete incident windows. \textbf{RCA accuracy} measures whether the generated diagnosis correctly identifies the annotated root cause, and it is computed as
\begin{equation}
    \mathrm{RCA}=\frac{1}{N}\sum_{i=1}^N\mathbb{1}\left[ \hat{f}_i=f_i \right],
\end{equation}
where $N$ is the number of root causes, and $\hat{f}_i$ and $f_i$ are predicted and ground-truth root-cause labels, respectively.
As shown in Fig.~\ref{fig:rq3_rca}, all methods degrade as semantic noise increases, but the ranking remains stable across all datasets and noise levels: \texttt{NeSy-Edge} performs best, followed by RAG-only and then Vanilla LLM. The improvement from Vanilla LLM to RAG-only indicates that historical references provide useful supplementary evidence, while the further gain of \texttt{NeSy-Edge} shows that retrieval alone is still insufficient under noisy conditions and that graph-derived causal evidence provides an additional and consistent benefit for root-cause diagnosis.

Specifically, HDFS starts from relatively strong low-noise accuracy as shown in Fig.~\ref{fig:rq3_rca}(a), but the two baselines drop much faster once the noise reaches the middle-to-high range, whereas \texttt{NeSy-Edge} remains comparatively stable and still achieves $0.70$ at noise $1.0$. From Fig.~\ref{fig:rq3_rca}(b), OpenStack shows the clearest advantage of structural evidence. Its control-plane incidents often contain short and semantically similar messages, so retrieval-only diagnosis can still be distracted by noisy local evidence; under maximum noise, \texttt{NeSy-Edge} retains an RCA accuracy of $0.76$, compared with $0.55$ for RAG-only and $0.42$ for Vanilla LLM. 
Hadoop is also challenging as depicted in Fig.~\ref{fig:rq3_rca}(c), as all methods show an earlier decline once noise is introduced; even so, \texttt{NeSy-Edge} degrades most gracefully, from $0.82$ at noise $0.0$ to $0.64$ at noise $1.0$, remaining above both baselines throughout. These results indicate that action-layer robustness begins with more reliable root-cause localization, which we further examine using the stricter E2E metric below.

\begin{figure}[!t]
    \centering
    \includegraphics[width=0.495\textwidth]{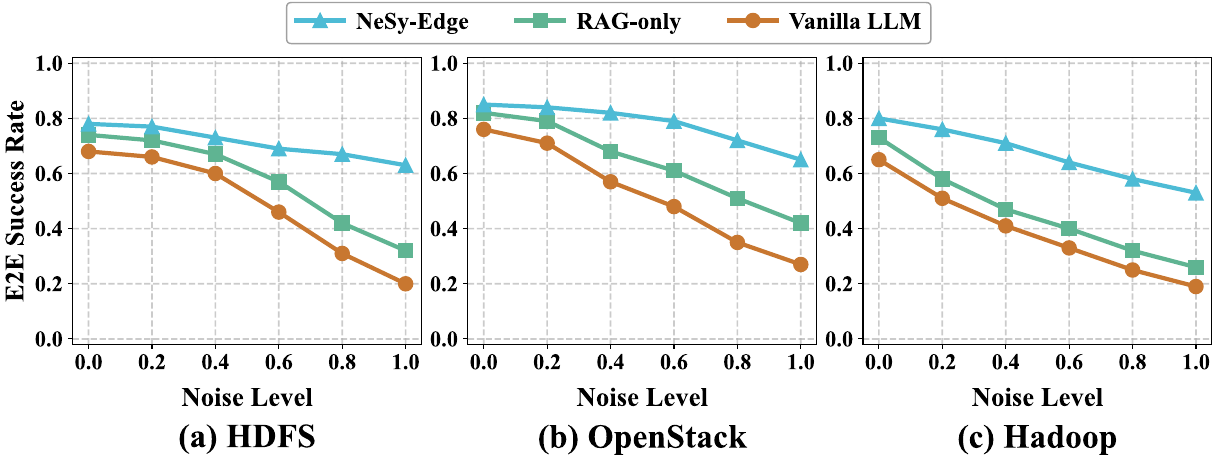}
    \caption{E2E success rate under increasing semantic noise.}
    \label{fig:rq3_e2e}
\end{figure}

% \subsubsection{End-to-End Success Rate (Action Layer)}
On the other side, \textbf{E2E success rate} is stricter than RCA accuracy because it requires both a correct diagnosis and a correct repair action. Accordingly, all curves in Fig.~\ref{fig:rq3_e2e} lie below their RCA counterparts. Nevertheless, the ordering remains unchanged across datasets and noise levels, with \texttt{NeSy-Edge} consistently outperforming RAG-only and Vanilla LLM. This shows that the benefits of historical references and graph-derived causal evidence do not stop at diagnosis quality, but carry over to more reliable end-to-end action decisions.

The gap between RCA and E2E is informative on Hadoop, where longer incident windows and stronger subtype confusion make it harder to translate a correct diagnosis into the right recovery action; at noise $1.0$, \texttt{NeSy-Edge} reaches $0.53$ E2E success, compared with $0.26$ for RAG-only and $0.19$ for Vanilla LLM. On HDFS, the two baselines experience a clear high-noise collapse in E2E, dropping to $0.32$ and $0.20$ at noise $1.0$, while \texttt{NeSy-Edge} still maintains $0.63$ as depicted in Fig.~\ref{fig:rq3_e2e}(a). From Fig.~\ref{fig:rq3_e2e}(b), OpenStack again shows the strongest end-to-end robustness and the clearest payoff of carrying graph-derived causal evidence into downstream action decisions: under maximum noise, \texttt{NeSy-Edge} achieves $0.65$, versus $0.42$ for RAG-only and $0.27$ for Vanilla LLM. Similarly, Hadoop also results in consistent performance under different noise levels as shown in Fig.~\ref{fig:rq3_e2e}(c). Overall, the action-layer results show that structured evidence from upstream perception and reasoning improves not only root-cause diagnosis, but also the stability of downstream repair decisions under noise.

\subsection{Discussion}
The performance superiority across the three layers suggests that the advantage of \texttt{NeSy-Edge} does not come from a single component, but from the complementarity of its neuro-symbolic design. In the perception layer, \texttt{NeSy-Edge} offers the strongest robustness to semantic noise while keeping latency far below that of a direct local SLM, demonstrating that cache matching, retrieval-assisted matching, and selective fallback together form a practical edge-side trade-off. In the reasoning layer, it produces more useful sparse causal graphs than the baselines, improving root-cause ranking while preserving interpretability. At the action layer, these upstream gains carry over to more stable RCA and E2E performance, indicating that graph-derived causal evidence complements historical retrieval for downstream diagnosis and repair action generation. Although the experiments are reported layer by layer to isolate individual contributions, the framework is deployed as an integrated pipeline, and the action-layer gains indicate that improvements in structured perception and causal reasoning are complementary and accumulate into stronger downstream robustness. Taken together, the layer-wise results show that robustness is built from perception to reasoning and finally to action. The consistent layer-wise improvements indicate that the end-to-end performance gain is not incidental, but arises from the contribution of each component in the overall framework. At the same time, this study still has limitations. First, the framework is primarily log-centric and does not yet incorporate multimodal runtime evidence such as traces, metrics, or topology signals, which could further improve fault understanding and recovery quality. Second, although \texttt{NeSy-Edge} follows an edge-first design, the final diagnosis stage may still depend on cloud reasoning.  These limitations are addressable and provide directions for future extension.

\section{Conclusion}\label{sec:concl}
This paper presented \texttt{NeSy-Edge}, a three-layer neuro-symbolic framework for trustworthy self-healing in the CC. The framework combines lightweight structured log parsing in the perception layer, prior-constrained causal discovery in the reasoning layer, and causal-evidence-guided diagnosis and recovery recommendation in the action layer. We used Loghub datasets (HDFS, OpenStack, and Hadoop) under increasing semantic noise. \texttt{NeSy-Edge} consistently outperformed layer-wise baselines, showing stronger robustness to semantic noise, lower edge-level latency than direct local SLM parsing, diagnostically useful causal graphs, and better downstream diagnosis and recovery performance. In particular, at the highest noise level, it retained parsing accuracy of 87\% on HDFS, over 93\% on OpenStack, and approximately 91\% on Hadoop; produced sparse causal graphs with 33, 39, and 41 edges and near-optimal AvgRank of 1.00, 1.74, and 1.28; and achieved up to 76\% RCA and 65\% E2E success while operating under resource-constrained edge nodes. These results show that combining symbolic constraints with selective neural inference is effective for edge-driven self-healing under noisy and heterogeneous conditions. Although our framework achieves strong performance, it is currently limited to log-centric self-healing and does not yet incorporate multimodal runtime evidence or continuous online adaptation. This limitation will be addressed in future work by extending the framework toward online self-healing for broader fault contexts and more adaptive recovery across the CC.

\bibliographystyle{IEEEtran}
\bibliography{ref.bib}
\balance
\end{document}